\documentclass[twocolumn,final,superscriptaddress]{revtex4-1}
\usepackage[utf8]{inputenc}
\usepackage{amsmath}
\usepackage{amsfonts}
\usepackage{amssymb}
\usepackage{graphicx}
\usepackage[ngerman,british]{babel}
\usepackage{color}
\let\oldsqrt\sqrt
\def\sqrt{\mathpalette\DHLhksqrt}
\def\DHLhksqrt#1#2{%
\setbox0=\hbox{$#1\oldsqrt{#2\,}$}\dimen0=\ht0
\advance\dimen0-0.2\ht0
\setbox2=\hbox{\vrule height\ht0 depth -\dimen0}%
{\box0\lower0.4pt\box2}}

\begin{document}
\title{Clogging at Pore Scale and Pressure Induced Erosion}

\author{R. Jäger} 
\email{jaegerr@ethz.ch} 
\affiliation{  ETH
  Z\"urich, Computational Physics for Engineering Materials, Institute
  for Building Materials, Wolfgang-Pauli-Strasse 27, HIT, CH-8093 Z\"urich
  (Switzerland)}

\author{M. Mendoza}
\email{mmendozaj@unal.edu.co}
\affiliation{  ETH
  Z\"urich, Computational Physics for Engineering Materials, Institute
  for Building Materials, Wolfgang-Pauli-Strasse 27, HIT, CH-8093 Z\"urich
  (Switzerland)}
\affiliation{
  Simulation of Physical Systems Group, Ceiba-Complejidad, Universidad Nacional de Colombia, Departamento de Fisica,\\
  Crr 30 \# 45-03, Ed. 404, Of. 348, Bogot\'a D.C., Colombia\\}
  
\author{H. J. Herrmann}
\email{hjherrmann@ethz.ch} 
\affiliation{  ETH
  Z\"urich, Computational Physics for Engineering Materials, Institute
  for Building Materials, Wolfgang-Pauli-Strasse 27, HIT, CH-8093 Z\"urich
  (Switzerland)}
\affiliation{Departamento de F\' isica, Universidade do Cear\' a, 60451-970 Fortaleza, Brazil}
\affiliation{on leave from C.N.R.S., UMR 7636, PMMH, ESPCI, 10 rue Vauquelin, 75231 Paris Cedex 05, France}

\begin{abstract}
Introducing a model to study deposition and erosion of single particles at microscopic scale, we investigate the clogging and erosive processes in a pore. The particle diameter, concentration, and adhesive forces rule the way particles are deposited, and therefore, characterize the clogging process. We study the hydraulic pressure that induces erosive bursts and conclude that this pressure depends linearly on the deposited volume and inversely on the pores' diameter. While cohesion does not play an important role for erosive bursts, the adhesion is the main force initiating clogging and when overcome by the hydraulic pressure, erosive bursts are triggered. Finally, we show how the magnitude of erosive bursts depends on the pore length, particle diameter and pore size.
\end{abstract}

\maketitle
\section{Introduction}
Erosion in porous media plays an important role in a variety of systems, for example sand production in oil reservoirs \cite{NAG:NAG154,ZHOU2011237} or breakthrough in water treatment plants. Bianchi \emph{et~al.} \cite{PhysRevLett.120.034503,Bianchi2018} recently discovered that critical bursts in filters follow a power-law and studied their statistical properties. With a theoretical model we investigated these erosive bursts using computer simulations \cite{PhysRevLett.119.124501} and found that they occur when the local fluid pressure overcomes the forces keeping deposited matter in place and blocked pathways get unclogged.
In this model we took several simplifying assumptions, namely, the motion of suspended particles was described by a concentration field, and thus deposits are continuous matter rather than a conglomerate of individual particles. While this model allowed us to study relatively large porous media, the deposition or erosion of single particles and the effect of particle size or pore size on the erosive behavior remained open questions. To answer these questions, studies at microscopic scale need to be performed.

While there are already studies of clogging and unclogging of microscopic channels, both experimental and theoretical, to our knowledge there is none yet that considers the phenomenon of erosive bursts. For example Sendekie \emph{et~al.} \cite{doi:10.1021/acs.langmuir.5b04218} studied the relation between hydrodynamic conditions and chemical properties of clogging, Agbangla \emph{et~al.} \cite{C4SM00869C} studied the effect of repulsive DLVO forces on clogging with simulations that couple computational fluid dynamics (CFD) with a discrete element method (DEM). Even though computationally expensive, coupled CFD-DEM have become very popular to tackle a variety of problems. Zhou \emph{et~al.} \cite{ZHOU2011237} for example used such a method to study liquid-induced erosion in weakly bonded sand, Lominé \emph{et~al.} \cite{NAG:NAG1109} used it to model piping erosion. 

Hence we investigate the erosive bursts on a smaller scale, where individual suspended particles are considered. These particles experience a short range cohesive force that can lead to clustering or flocking and an adhesive force that is responsible for the deposition on a solid surface. We model a solid pore using voxels that define the fluid domain boundary and interact with suspended particles.
Our model is based on the one by Lominé \emph{et~al.} and as they do, we use a lattice Boltzmann method (LBM) to resolve the fluid flow through the pore space and to calculate the fluid drag exerted on each suspended particle. We also use a discrete element method (DEM) and a dash-dot force to calculate the interaction between particles. However we implemented cohesive forces that can dynamically attach and detach particles depending on the balance of cohesive and other forces. The cohesive force has a similar form as proposed by Zhou \emph{et~al.} \cite{ZHOU2011237}, though cohesive bonds can be formed and broken dynamically. Also we model an adhesive force acting between suspended particles and pore surface. The interaction between particles and pore surface also features a Coulomb friction force, that restricts deposited particles from sliding along the surface. With these ingredients we are not only able to model the deposition and erosion of single particles or clusters of particles, but also investigate the dependencies of pore and particle specifics on erosive bursts. 
We have found that cohesion plays a minor role and the adhesion is the decisive force leading to erosive bursts. Finally we find that the pressure required to initiate an erosive burst is linearly dependent on the deposition length, which we define by the deposited volume divided by the pore area. Thus showing that our previous assumption that a critical pressure gradient has to be overcome to initiate an erosive burst \cite{PhysRevLett.119.124501} seems indeed reasonable. 

This paper is organized as follows: in section \ref{sec:theory} the model will be described in detail, section \ref{sec:results} will show the results and in section \ref{sec:conclusions} conclusions will be drawn and an outlook given.

\begin{figure}[h!]
\includegraphics[width=\columnwidth]{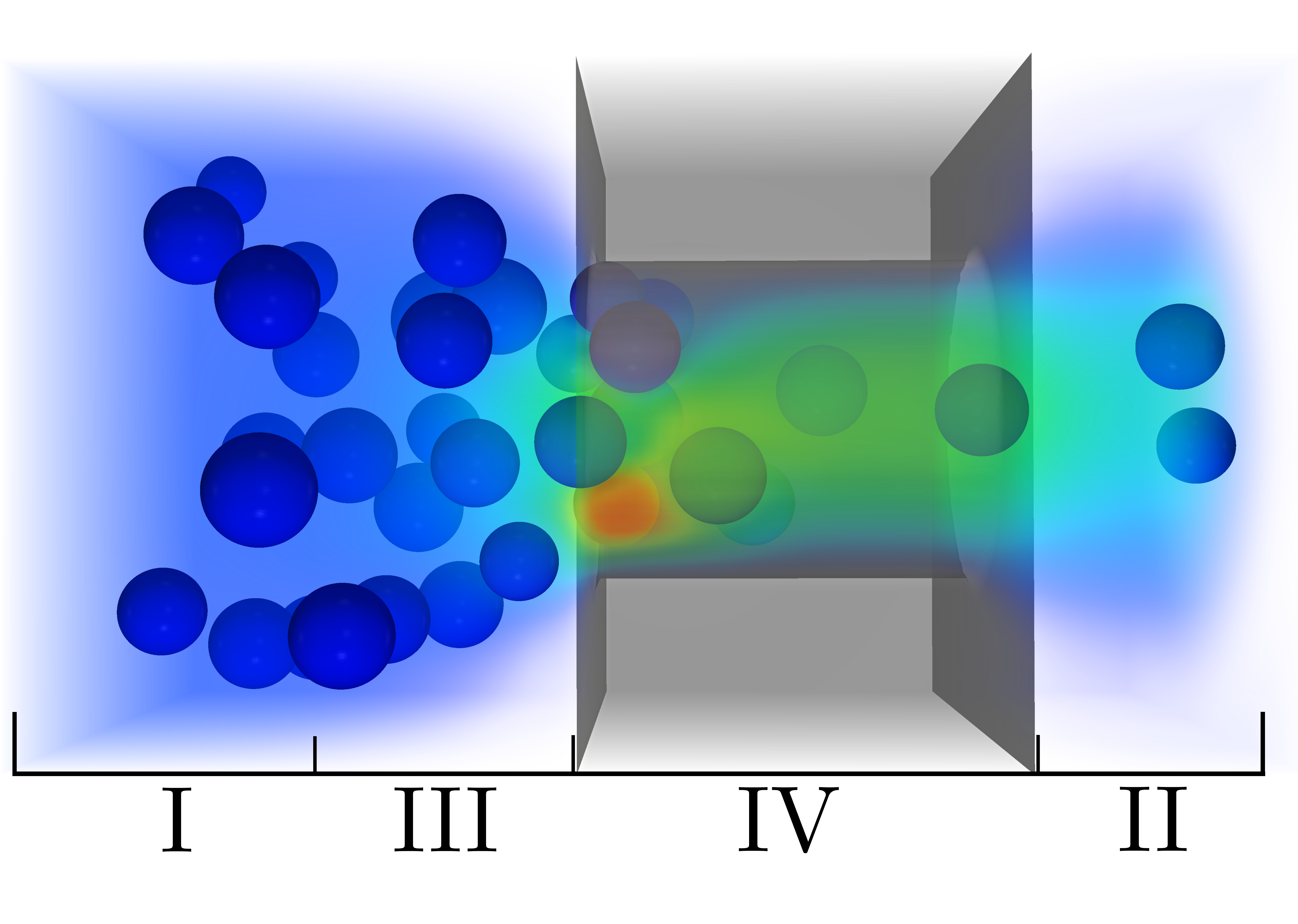}
\caption{Setup for the simulations, in the inlet region (I) particles are inserted until a certain inlet particle concentration $C_{\text{in}} \sim C_0$ is reached. When particles leave the outlet region (II) they are disregarded. The particles flow freely through region (III) and reach the pore structure (IV) which in this case is a simple constriction with the shape of a cylinder.}
\label{fig:setup}
\end{figure} 

\section{Model Description}
\label{sec:theory}
The purpose of this study is to simulate erosion and deposition at a microscopic scale, where individual particles are entrained by a fluid. These suspended particles experience a drag force from the fluid and in turn influence the fluid via moving boundary conditions. The particles also have repulsive forces that restrict overlap and short range cohesive forces that pull them together if two of them are close to each other. Furthermore, there is a static solid matrix in the shape of a pore through which the fluid is driven, and particles that are close to its surface experience an adhesive force pulling them towards the surface. Coulomb friction acts on particles that are in contact with the pores' surface.

To describe the fluid and particle dynamics we base our model on the one proposed by Lominé \emph{et~al.} \cite{NAG:NAG1109}, where the lattice Boltzmann method is used to calculate the fluid flow and a discrete element method is used for the particle dynamics. The structure of the three-dimensional pore is static and described using voxels, such that the interaction between particles and wall is different from the particle-particle interaction. Furthermore, we incorporate a simplified cohesive force by introducing a short range force that acts when particles are close together. The same is done when particles approach the wall, this adhesive force has the same form but can have a different amplitude than the cohesive force. Hence we investigate cohesive and adhesive forces and their effect on deposition and erosion. Note however that we consider suspended particles which are of micrometer scale and have very short range and weak attractive forces. These can be Van der Waals forces or other attractive forces, for instance contact friction due to the roughness of particles and pore surface. 

\subsection{Fluid Dynamics}
To solve the Navier-Stokes equations we use the lattice Boltzmann method (LBM) (see \cite{wolf2000lattice}). The lattice Boltzmann equation describes the dynamics of discrete distribution functions $f_i$ that discretize fluid density and velocity fields at each point of a lattice:

\begin{equation}
\label{eq:lbm}
		f_i(\mathbf{x}+\mathbf{c}_i, t+\delta t ) - f_i(\mathbf{x},t) = \Omega(\mathbf{f}),
\end{equation}
where $\Omega$ is the two-relaxation-time collision operator as shown by Talon \emph{et~al.} \cite{talon2012assessment}. The fluid is driven by a constant inlet fluid velocity implemented as described in Ref. \cite{1742-5468-2009-06-P06015} or by a constant pressure loss between in- and outlet.

To solve the boundary conditions for the LBM more accurately we employ an interpolation scheme developed by Mei \emph{et~al.} \cite{PhysRevE.65.041203} where bounced back distribution functions are corrected according to the distance and the velocity of the boundary. The hydrodynamic drag force exerted on a suspended particle is calculated as follows (see Ref. \cite{PhysRevE.65.041203,NAG:NAG1109}):

\begin{equation}
\label{eq:dragforcecalc}
\mathbf{F}_h = \frac{1}{\delta t} \sum\limits_i \left( 2 f^*_i - 6 \rho \omega_i \mathbf{u}_w \cdot \mathbf{c}_i \right) \mathbf{c}_i ,
\end{equation}
where the sum runs over all distribution functions that are bounced back from a particle wall whose velocity is denoted by $\mathbf{u}_w$. The distribution functions are corrected by the second term in the sum of Eq. \eqref{eq:dragforcecalc}, thus the equation simply sums all momenta exchanged between fluid nodes and particle. In turn, the LBM takes into account the position and velocity of the particles' wall. To accomplish this the intersection points of lattice Boltzmann vectors and the spheres surface must be determined, this boundary problem is solved as described in Ref. \cite{IGLBERGER20081461}.

\subsubsection{Lubrication Force}
When two particles come very close together, such that there are no fluid nodes between them anymore, the hydrodynamic force calculated in Eq. \eqref{eq:dragforcecalc} becomes very inaccurate as it considers void of fluid nodes as vacuum. Therefore we correct the hydrodynamic force acting on a particle in this case using the lubrication force (see Ref. \cite{ZHANG200592})
\begin{equation}
F_\text{lub} = 6 \pi \rho \nu a^2 v / h ,
\end{equation}
where $v$ is the relative velocity between particles, $h$ the distance between them and $a = R_i R_j / (R_i + R_j)$, where $R_i$ and $R_j$ are the radii of the two particles. This lubrication force diverges for $h \to 0$ and is not appropriate for extremely short separations, thus we introduce a cutoff at $h \sim 0.01 R$ which can be associated with the roughness of the particles. 

\subsection{Particle Dynamics}
\subsubsection{Particle-Particle Interaction}
Particles in our model are simple spheres and the Verlet integration algorithm, standardly used in DEM, is used to calculate their motion. The particle-particle interactions are described using the dash-dot model \cite{delapproach}. The repulsive force in this model is described by a stiffness $k_n$ which acts as a spring if two particles overlap and for numerical stability a damping term is added with damping coefficient $k_d$: 

\begin{equation}
\label{eq:dashdot}
F_n = k_{n} \delta + k_d \dot{\delta} ,
\end{equation}
$\delta$ is the positive overlap. The cohesion is modeled with an additional term if the distance between particles is very small, smaller than a fraction ($h \leq 0.03 (R_i + R_j)$) of the radii of involved particles:

\begin{equation}
\label{eq:cohesiveforce}
F_c = h k_c \frac{R_i R_j}{R_i + R_j}, 
\end{equation}
where $F_c$ is the cohesive force between two particles denoted by $i$ and $j$. The cohesive force depends on the radii of the particles which is motivated by the van der Waals forces between two spheres \cite{Israelachvili2011253}. A similar cohesive force was used by Zhou \emph{et~al.} \cite{ZHOU2011237}. Note that we are only modeling a very simplified cohesive force of short range whose amplitude can be tuned by changing the cohesion coefficient $k_c$.
To calculate the total force from particle-particle interactions, equations \eqref{eq:dashdot} and \eqref{eq:cohesiveforce} have to be added up for all interacting particles:

\begin{equation}
\mathbf{F}_\text{pp} = \sum\limits_j \left( F_{n,j} + F_{c,j} \right) \mathbf{\hat{n}}_j .
\end{equation}

\subsubsection{Particle-Wall Interaction}
The pore structure is modeled using voxels that are placed on the same grid as used by the LBM for the fluid solver. The voxels have a mass index between 0 and 1, for mass index 0 it is a fluid node and for one a solid cell (see Ref. \cite{PhysRevE.95.013110}). Using this scheme we can construct any pore shape desired. Thus we treat the particle-wall interaction in the same way as the particle-particle interaction where voxels behave like very large particles and are static. The dash-dot model shown in Eq. \eqref{eq:dashdot} is used again with a different stiffness for the repulsive force and there is an adhesive force analogous to the cohesive force:
\begin{equation}
\label{eq:adhesiveforce}
F_a = h k_a R_i ,
\end{equation}
where $h$ is the distance between the particle and the wall, $k_a$ is the adhesion coefficient and $R_i$ the particle radius. We use the same cutoff as for the cohesive force ($h \leq 0.03 R_i$). The dependence on the radius of the particle is here motivated by the van der Waals force between a sphere and a wall \cite{Israelachvili2011253}. A particle adhering to a wall could slide in tangential direction if only the aforementioned adhesive and the repulsive forces are considered. Therefore a frictional force is introduced that hinders the particle from moving in the tangential direction as long as the particle adheres to the pore wall. This friction is a simple dynamic Coulomb friction which is proportional to the normal force between particle and wall:
\begin{equation}
F_f \leq \mu F_n ,
\end{equation}
with a constant friction coefficient $\mu$, acting in the tangential direction. The total particle-wall force is calculated by summing up all the interactions between particle and wall-voxels:
\begin{equation}
\mathbf{F}_\text{pw} = \sum\limits_k \left( F_{n,k} + F_{a,k} \right) \mathbf{\hat{n}}_k + F_{f,k} \mathbf{\hat{t}}_k  , 
\end{equation}
where $k$ runs over all voxels interacting with the particle. Note that this sum usually does not include very many voxels as the adhesion is a short range force and the normal force from the dash-dot model prevents large overlaps.

\subsection{Coupling of Fluid and Particles}

Since we consider small sized suspended particles we can neglect the rotation of the particles since the rotational energy is much smaller than the kinetic energy. The total force acting on a particle is then the sum of all interactions:
\begin{equation}
\mathbf{F} = \mathbf{F}_\text{pp} + \mathbf{F}_\text{pw} + \mathbf{F}_h ,
\end{equation}
and the motion of particles simply follows Newtonian mechanics: 
\begin{equation}
\mathbf{\ddot{x}} = \frac{\mathbf{F}}{M} .
\end{equation}
To calculate the trajectories of particles the Verlet algorithm is employed, unfortunately choosing the same time step for the LBM and the DEM is either extremely slow or unstable. Thus we adopt the scheme from Lominé \emph{et~al.} \cite{NAG:NAG1109}, where several DEM time steps are calculated between steps of the LBM: 
\begin{equation}
\delta t _\text{DEM} = \frac{\delta t}{ N_\text{DEM} }.
\end{equation}
The time steps for the DEM has to be small enough for stability reasons, the higher the stiffness coefficient in Eq. \eqref{eq:dashdot} the smaller the DEM time step needs to be. When the two solvers are coupled, the DEM takes into account a fraction of the hydrodynamic force in each time step, the fluid solver only requires the position and velocity of the particles. The setup of a typical simulation is depicted in Fig. \ref{fig:setup}, where the inlet boundary is at the left side of region I and the outlet boundary at the right side of II. The surface of the pore structure is shown in region IV. In the inlet region (I) particles are introduced with a diameter drawn randomly from a Gaussian distribution of an average diameter $d = 1$ and a standard deviation of $0.1 d$. While the diameters do not vary very much, this should prevent crystalline structures of clustered particles. We define the diameter to be one and scale all other lengths accordingly. The particles are placed with a minimum distance of $0.25 d$ to each other. 

\section{Results}\label{sec:results}
\begin{figure}[h!]
\includegraphics[width=\columnwidth]{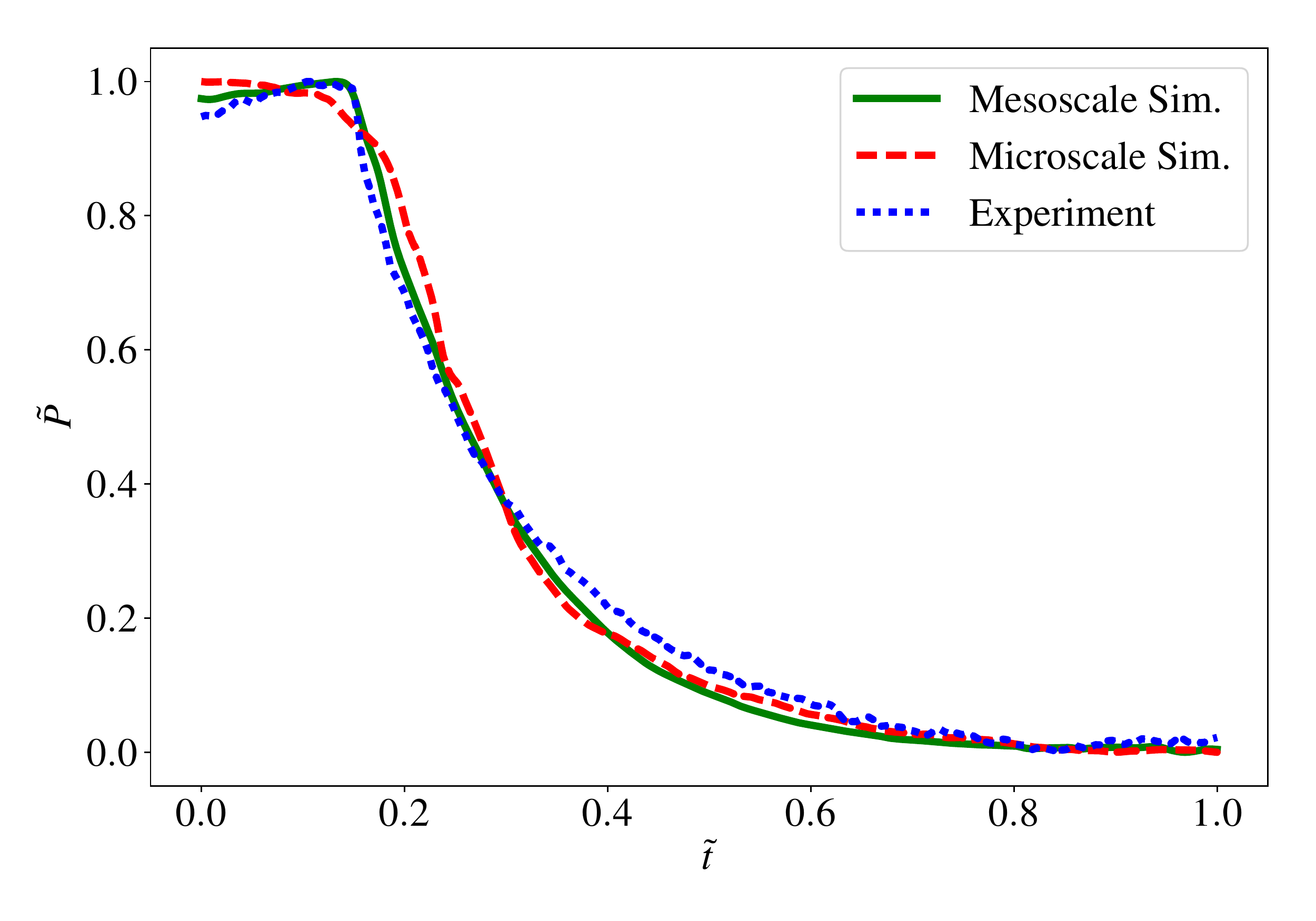}
\caption{The experimental data stems from Ref.  \cite{PhysRevLett.120.034503} and shows a jump in pressure loss caused by an erosive burst. Data from our previous model reproducing erosive bursts are shown by the continuous line and data from the here presented model is shown by the dashed red line. Pressure and time are rescaled for all three curves by the respective jump size ($\tilde{P} = P / \Delta P$) and jump duration ($\tilde{t} = t / \Delta t$).}
\label{fig:singlejump}
\end{figure}

In our first set of simulations we set a constant fluid velocity at the inlet boundary.
Recently we introduced a model that was able to reproduce and explain erosive bursts in porous media \cite{PhysRevLett.119.124501}. In this model particle flow is approximated using the convection-diffusion equation. Here we model the deposition and erosion using more fundamental principles. When the adhesive force is strong enough such that particles deposit inside the pore space, the pressure through the pore increases. When a certain pressure is reached the hydrodynamic forces exceed the adhesive forces and the deposited particles get re-entrained. This is accompanied by a sudden reduction in fluid pressure loss through the pore model. To verify that we are indeed finding the same erosive behavior we compare this jump in pressure with the previous model and the experimental data from Ref. \cite{PhysRevLett.120.034503} in figure \ref{fig:singlejump}. And indeed we find that the shape of the pressure jump in the current simulation, which we call the microscale simulation agrees with the previous model and the experimental data. The curve from the mesoscale simulation however more closely resembles the experimental measurement, because the considered size of the porous medium and number of particles detached is much larger than in the present microscopic approach. Bianchi \emph{et~al.} \cite{Bianchi2018} have shown that often whole clusters of pores are re-opened by erosive bursts. While in our previous model \cite{PhysRevLett.119.124501} a small part of a filter was simulated, in the present study the system only considers a single pore which we show to be the minimal system able to exhibit erosive bursts. Also it is worth noting that while here the pressure loss is measured on one pore, in the mesoscopic systems it was measured for a whole porous structure including many pores.

\begin{figure}[h!]
\includegraphics[width=\columnwidth]{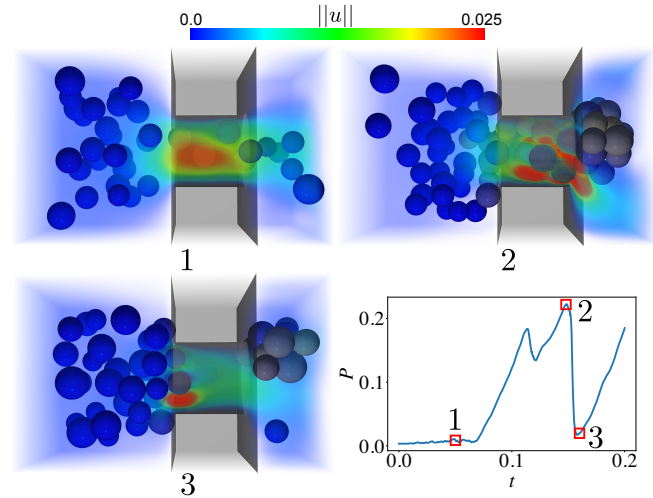}
\caption{Three snapshots of a simulation are shown, the numbers of the snapshots relate to time and pressure shown in the graph. The first snapshot (1) shows a state before any deposition has taken place, in the second (2) the deposition and pressure loss has reached a local maximum (peak pressure) and in the third (3) an erosive burst has unclogged the pore space. The particle colors from blue to yellow indicates the experienced fluid drag per mass for each particle (from low to high). The rainbow colors indicate the flow speed of the fluid.}
\label{fig:collection}
\end{figure} 

\subsection{Critical Erosive Pressure}
One goal when developing this model for erosion and deposition at microscopic scale was to study if we could define a conclusive criterion for the pressure that initiates an erosive burst. In our previous model \cite{PhysRevLett.119.124501} we assumed that this critical pressure is proportional to the width of the deposit that has to be dislodged 
\begin{figure}[h!]
\includegraphics[width=\columnwidth]{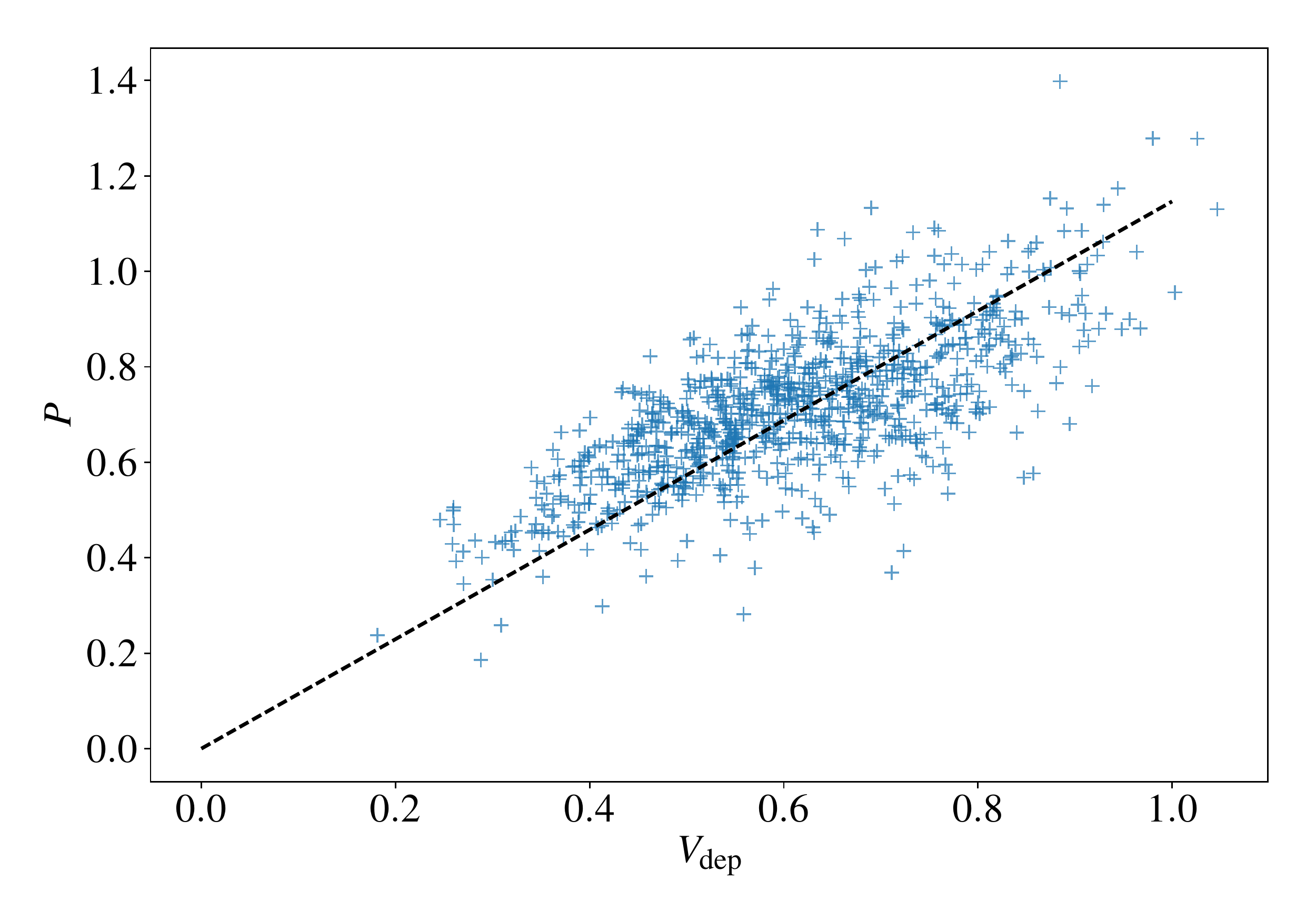}
\caption{Pressure loss peaks $P$ are plotted versus the deposited volume $V_\text{dep}$ from 100 simulations with different random seeds. The dashed line is a linear fit with slope $1.15(1)$ and the Pearson correlation coefficient $0.71(1)$ suggests positive correlation.  }
\label{fig:dropvol}
\end{figure} 
\begin{equation}
	P_\mathrm{crit} = L_\mathrm{dep} \sigma_c ,
\end{equation} 
where $L_\mathrm{dep}$ is the width of the deposit and $\sigma_c$ is the threshold for erosion. As we here have a single pore, we can measure the pressure on both sides of the pore easily. We can also measure the deposited volume $V_\mathrm{dep}$, which we measure by summing up all volumes of particles that are deposited and located inside the pore radius. Deposited particles are particles that are in a cohesive or adhesive bond and have zero or very low velocity, such that moving flocks of particles are not counted as deposited particles. In figure \ref{fig:dropvol} we see that there indeed seems to be a linear relation between critical pressure and deposited volume. 

\begin{figure}[h!]
\includegraphics[width=\columnwidth]{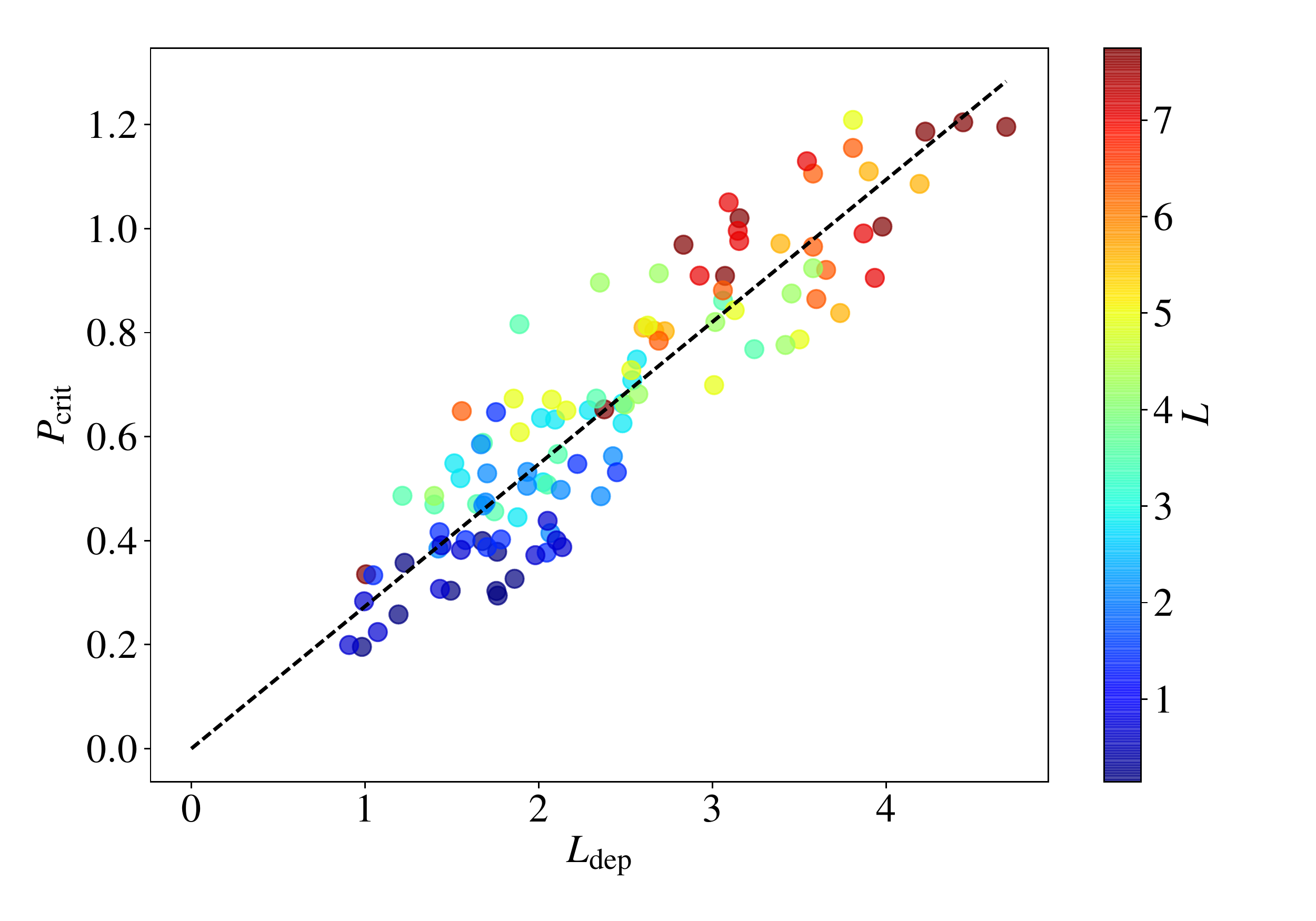}
\caption{Twelve simulations were run for different pore lengths $L = \{ 0.14, ..., 7.75 \}$ (see \ref{app:porelength}) and constant pore diameter $D = 3.52$. The lengths are given in dimensionless units by dividing by the particle diameter $d$. The pressure peaks show an approximate linear correlation against the deposited length $L_\text{dep}$. The Pearson correlation coefficient is $0.91(1)$. The dashed line is a linear fit over all data points with coefficient $0.27(1)$.  }
\label{fig:porelength}
\end{figure}

The width of the deposit can be estimated by dividing the deposited volume by the pore cross section area

\begin{equation}
	L_\mathrm{dep} = \frac{V_\mathrm{dep}}{D^2 \pi / 4}. 
\end{equation} 
Now we investigate whether this relation holds if we change the pore geometry, first we change the length of the pore and check if the relation between the critical pressure, inducing an erosive burst, and the deposition width is still the same. In figure \ref{fig:porelength} we show a set of simulations where the color shows to what pore length a specific point belongs.

\begin{figure}[h!]
\includegraphics[width=\columnwidth]{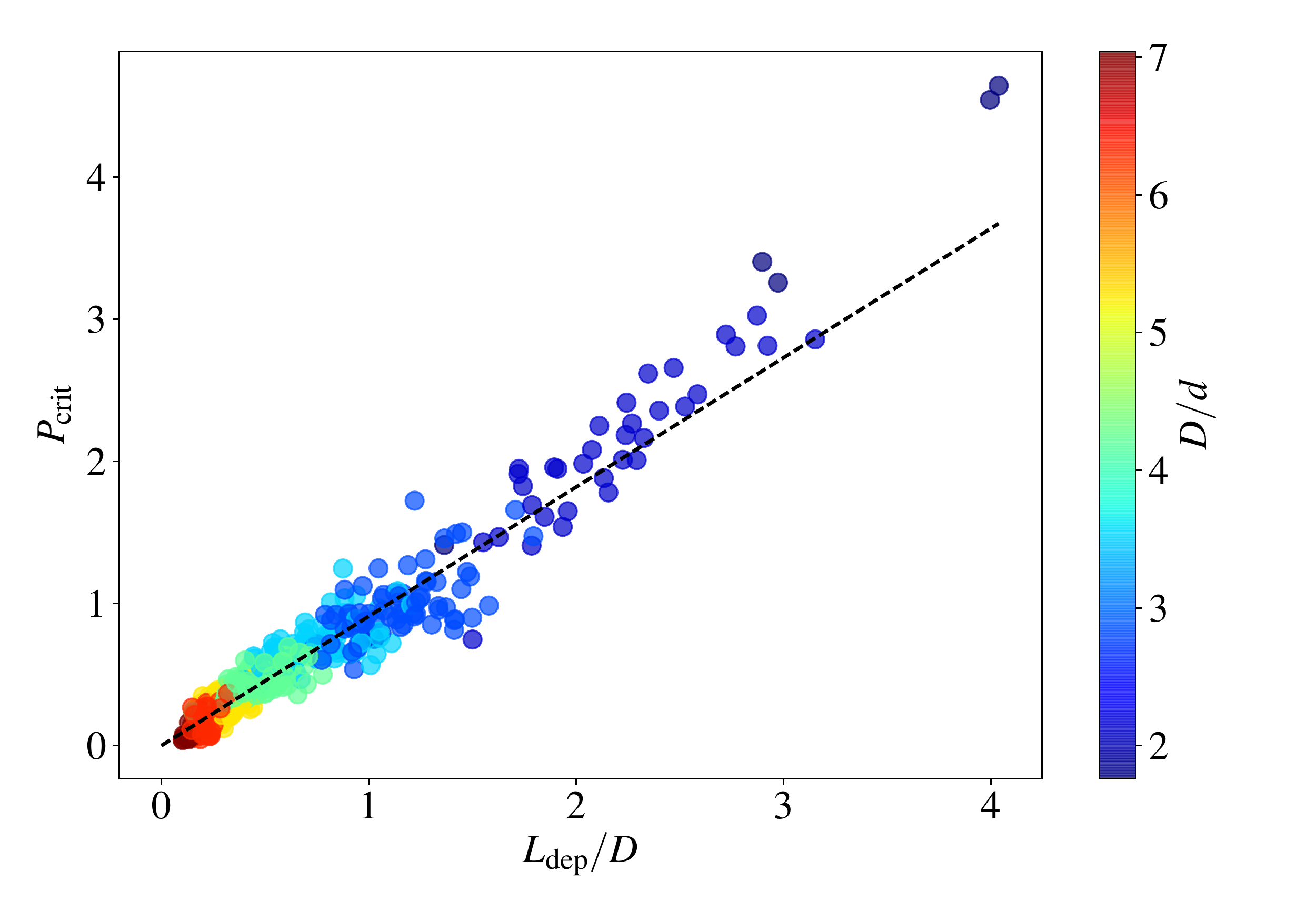}
\caption{A set of simulations was run for different pore diameters $D= \{ 1.76, ..., 7.04 \}$ (see \ref{app:porediameter}) and constant pore length $L = 4.22$. The pressure peaks are plotted against $L_\text{dep} / D$. Smaller pore diameters exhibit larger critical pressures, the same adhesive coefficient was used for all simulations. The dashed line is the graph of equation \eqref{eq:pfit}.}
\label{fig:porediameter}
\end{figure}

While on average longer pores lead to a higher critical pressure, the relation between pressure and deposited width is constant. Second we change the pore diameter, while keeping everything else the same, including the ratio between inlet area and pore cross section area. We found that for a given pore diameter the points $(P,L_\mathrm{dep})$ are linearly correlated, however they do not fall onto the same line. Thus we calculated the slope for all of them and, using an exponential fit, found that the more generic relation between critical pressure and deposited width is 
\begin{align}
\label{eq:pfit}
P_\mathrm{crit} = K L_\mathrm{dep} D^{-1} , \qquad &K = 0.91 \pm 0.04 .
\end{align}
Thus our simulations indicate that the critical pressure is inverse proportional to the pore diameter. In figure \ref{fig:porediameter} we show the simulation results for different pore diameters and rescale the $x$-axis according to Eq. \eqref{eq:pfit}.

\subsection{Theoretical analysis of the critical erosive pressure}
We found that one can estimate the critical erosive pressure when two simplifying assumptions are made. First we assume the deposited particles are packed tightly in a cylindrical shape, and second that there is a mean adhesive strength between particles and wall. The following theoretical analysis shows where the numerically found relation between critical erosive pressure and pore diameter $P \propto L_\mathrm{dep} / D$ comes from. We know that an erosive burst occurs when hydrodynamic forces overcome the adhesive forces. The total hydrodynamic force can be estimated as the pressure loss times the pore cross section area
\begin{equation}
	F_H \sim P \frac{D^2 \pi}{4}. 
\end{equation}
Now just as for the deposited width we assume that all deposited particles are packed tightly together inside the pore, furthermore we assume that there is a mean value for the adhesive force $f_a$ per area between particles and surface. The total adhesive force is then approximated by an integral over the surface that is shared by the deposited particles and the pores' surface:
\begin{equation}
	F_A = \sum_k F_{a,k} \sim \int\limits_\Sigma f_a = f_a \Sigma = f_a \pi D L_\mathrm{dep} .
\end{equation}
Just before an erosive burst, the drag forces are equal to the adhesive forces, thus we can write the balance equation:

\begin{align}
F_H & \stackrel{!}{=} F_A \qquad &\Rightarrow \\
P \frac{D^2 \pi}{4} & \stackrel{!}{=} f_a \pi D L_\mathrm{dep} \qquad &\Rightarrow \\
P &= \frac{4 f_a L_\mathrm{dep}}{D}
\end{align}

or as function of the deposited volume
\begin{equation}
\label{eq:pvdep}
P = \frac{16 f_a V_\mathrm{dep}}{\pi D^3} .
\end{equation}
Using a mean field approximation for adhesive forces, our analysis shows that the relation between critical pressure and pore diameter follows naturally. While the exact pressure that induces an erosive burst depends on the specific configuration of deposited particles and pore geometry, an analytically derived estimate seems to serve well for qualitative purposes. 

\subsection{Parameter Phase Space}
\begin{figure}[h!]
\includegraphics[width=\columnwidth]{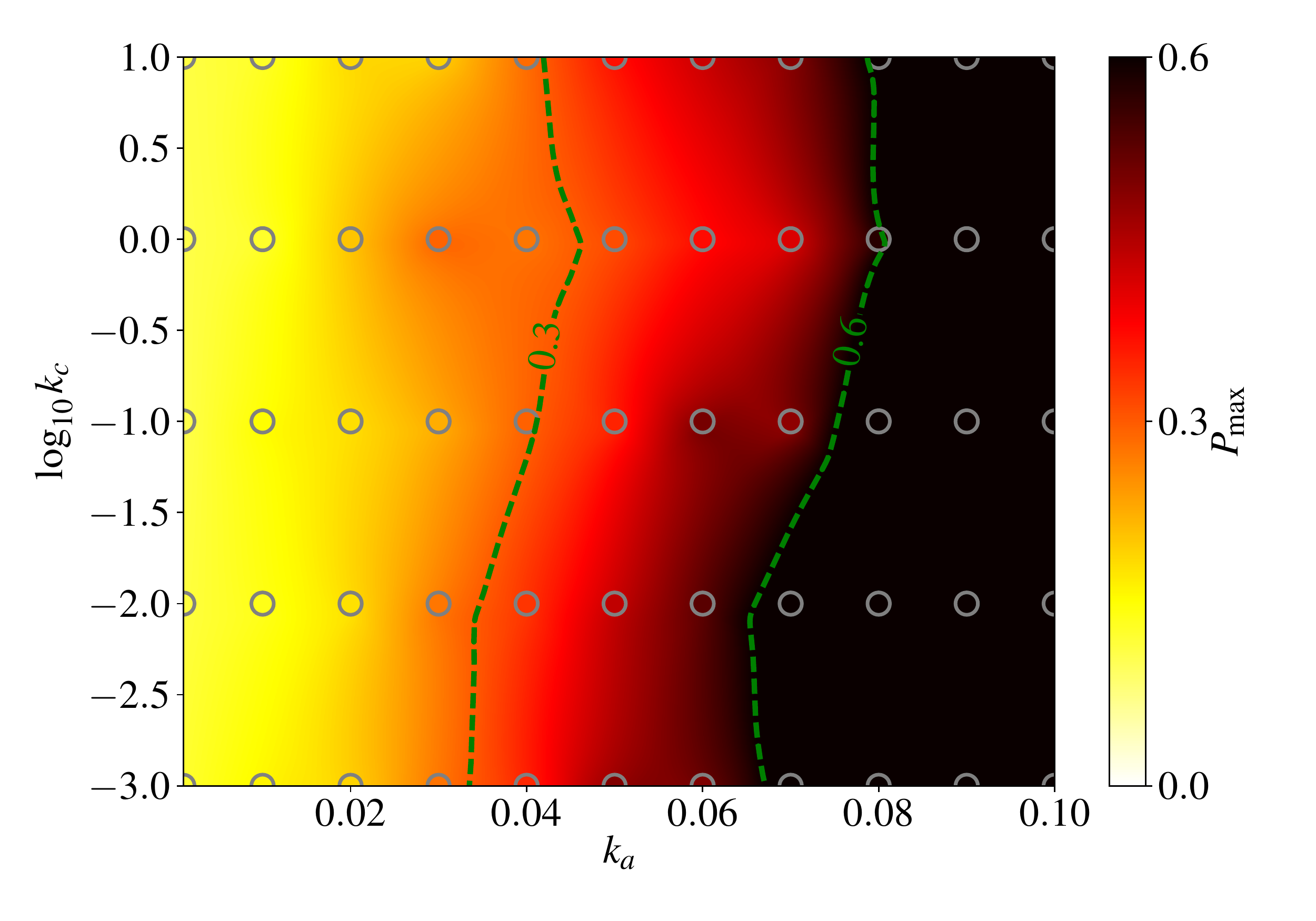}
\caption{Parameter phase space diagram for adhesive $k_a$ and cohesive $k_c$ coefficients. The circles represent data from simulations, the data in between is interpolated. The pore length is $L=6.45$, the pore diameter $D=3.52$.}
\label{fig:phaseac}
\end{figure}

\begin{figure}[h!]
\includegraphics[width=\columnwidth]{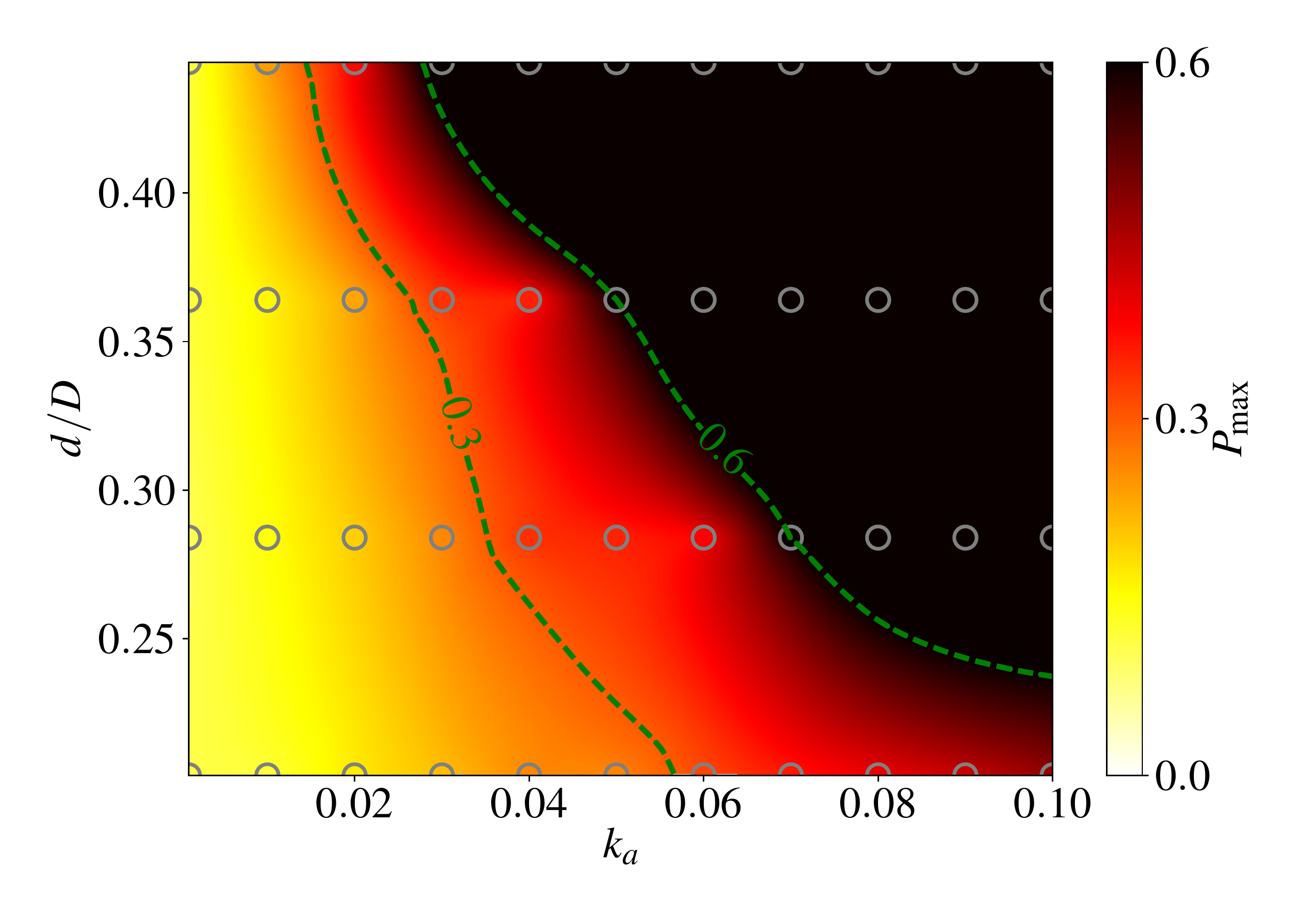}
\caption{Parameter phase space diagram for adhesive coefficient $k_a$ and particle diameter $d$. The circles represent data from simulations, the data in between is interpolated. Since we change here the particle diameter, lengths are given in units of the pore diameter $D$, the pore length is $L=1.2[D]$.} 
\label{fig:phasead}
\end{figure}

\begin{figure}[h!]
\includegraphics[width=\columnwidth]{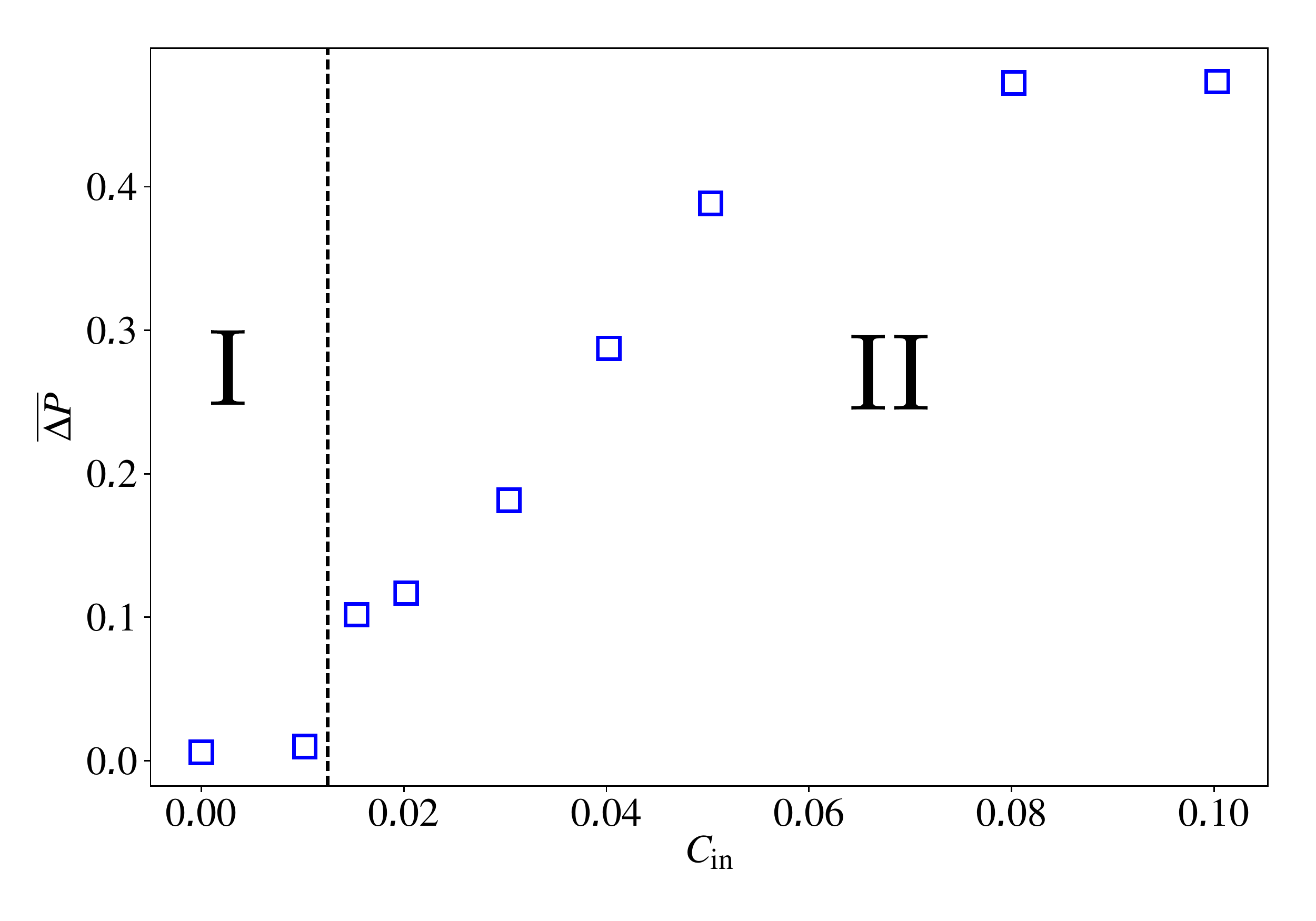}
\caption{At the inlet different particle concentrations are set $C_\text{in} = {0,...,0.1}$. The average pressure loss jump is measured for all simulations. The pore length is $L=4.23d$ and the pore diameter $D=3.52d$. For low concentrations (range I) there is no clogging, while for higher concentrations (range II) there is severe clogging.} 
\label{fig:conc}
\end{figure}

In this section we present our studies concerning the parameter phase space. First we investigated how the cohesive and adhesive forces influence the clogging behavior. We ran a large set of simulations where only the coefficients of these forces were varied, the results are shown in figure \ref{fig:phaseac}.
When the maximum pressure loss rises above $P_\text{max}[P_\text{out}] > 0.3$, the pore can be considered slightly clogged, and for $P_\text{max}[P_\text{out}] > 0.6$ there is severe clogging (see the dashed contour lines in Fig. \ref{fig:phaseac}). The $x$-axis is in linear scale and one can see that changing the adhesion within one order of magnitude changes the deposition behavior completely, from no deposition to severe clogging. On the other hand the cohesion coefficient was changed over several orders of magnitude but no significant change in the clogging behavior was observed. Going further we studied how the ratio between particle and pore diameter $d/D$ changes the clogging behavior. Again we consider a pressure loss above $0.3P_\text{out}$ as slight clogging, and above $0.6 P_\text{out}$ as severe clogging (see the dashed contour lines in Fig. \ref{fig:phasead}). The diagram in figure \ref{fig:phasead} shows the maximum pressure loss reached in the simulations. The simulations show that larger particles lead to larger pressure build-up and consequently to larger erosive bursts.
Another very important parameter is the inlet particle concentration $C_\text{in}$. Experiments (see Ref. \cite{PhysRevLett.120.034503}) have shown that the particle concentration is a crucial factor for the clogging and erosive behavior in deep bed filtration. Therefore we run a set of simulations where only $C_\text{in}$ is varied. The result is shown in figure \ref{fig:conc} and shows that while for low concentrations there is no clogging, above a certain concentration (dashed line at $C_\text{in} \sim 0.013$) there is clogging and consequently erosive bursts, which can be seen by the average jump in pressure loss $\overline{\Delta P}$. This result is in qualitative agreement with the observations shown in the experimental work by Bianchi \emph{et~al.} \cite{PhysRevLett.120.034503}.

\subsection{Constant pressure loss}
\begin{figure}[h!]
\includegraphics[width=\columnwidth]{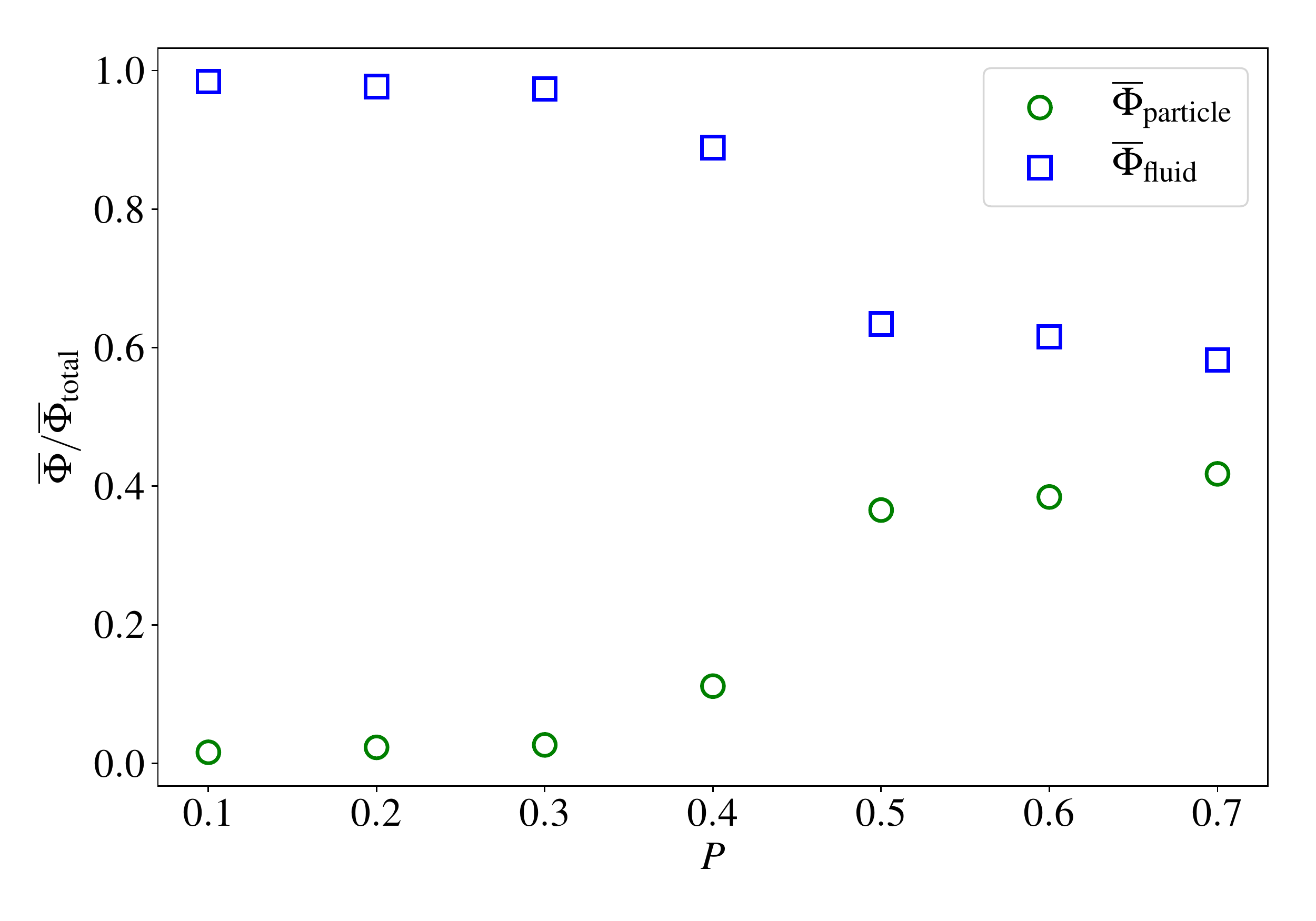}
\caption{For a constant pressure loss we measure average particle $\overline{\Phi}_\text{particle}$ and fluid flux $\overline{\Phi}_\text{fluid}$, normalized by the total flux. Up to a certain pressure ($P \sim 0.35$) particles are only deposited and there is no particle flux through the pore, above this pressure there is erosion of particles and a finite particle flux is measured at the outlet. The adhesion coefficient is $k_a = 0.05$.} 
\label{fig:pvpflux}
\end{figure}

In this section we present our results for a constant pressure loss, imposed between in- and outlet. Since the deposited particles are not packed densely enough to have total clogging, there is always a small flux of fluid remaining in this scenario. However we see in figure \ref{fig:pvpflux} that below a certain pressure loss there are no particles passing through the pore. Above this threshold the particle flux increases rapidly.

\section{Conclusion and Outlook}\label{sec:conclusions}
We presented a new model to investigate erosion and deposition at microscopic scale. We use simple adhesive and cohesive forces and study their effect on the clogging and erosive behavior inside a pore. While the adhesion proved crucial for clogging, the cohesion showed little influence. Note that this might only be true for particles close to pore size and short range forces. Our results confirm our previous finding \cite{PhysRevLett.119.124501} that the hydraulic pressure gradient acting on deposit is an erosive mechanism that can lead to erosive bursts. The critical pressure gradient causing an erosive burst was studied and we found that it depends strongly on the particle size and pore length. Furthermore we found that it depends linearly on the deposited volume and is inverse proportional to the pore diameter cube (see Eq. \eqref{eq:pvdep}).
A further step would be to have a more complex model for the cohesive and adhesive forces such as the simplified Johnson-Kendall-Roberts (JKR) model \cite{delapproach} or use non-spherical particles. 
Since the deposit itself can be seen as a changing porous medium, it would also be interesting to study the compaction and porosity of the deposit depending on the hydraulic pressure. Also more complex pore shapes could be studied which is easily doable with our model. Instead of just having one pore one could extend the static matrix to several pores or a full porous medium, however for this the code would need to be up-scaled.

\begin{acknowledgments}
We thank P. Iliev for the helpful input concerning discrete element methods. 
We acknowledge financial support from the European Research Council (ERC) Advanced Grant 319968-FlowCCS.
\end{acknowledgments}

\appendix
\section{Appendix}\label{sec:app}

\subsection{Model Parameters}
\label{app:modelpara}
Table \ref{tab:para} shows the model parameters used that are kept the same throughout all simulations, if not explicitly stated otherwise. Many of the parameters are only relevant for stability and performance. The damping coefficients are chosen to be half of the critical damping. Since we only consider laminar flow we chose a fast relaxation time ($\tau = 1$) to speed up the simulations, which for LBM gives a fluid viscosity of $1/6$. The ratio between fluid and particle density is chosen to be $2.3$, since this is the approximate ratio between the density of silica and water which was used in the experiments of Bianchi \emph{et~al.} \cite{PhysRevLett.120.034503,Bianchi2018}.

\begin{table}[h!]
\label{tab:para}
\begin{center}
  \begin{tabular}{ | c | c | c | }
    \hline
    description & symbol & value \\ \hline
    Fluid density & $\rho$ & 1 \\ \hline
    Particle mass density & $\rho_p$ & 2.3 \\ \hline
    Fluid viscosity & $\nu$ & $1/6$ \\ \hline
    Number of DEM steps per LBM step & $N_\text{DEM}$ & 100 \\ \hline
    Number of time steps per simulation & $T / \delta t$ & $10 ^6$ \\ \hline
    Cohesive coefficient & $k_c$ & 0.1 \\ \hline
    Adhesive coefficient & $k_a$ & 0.1 \\ \hline
    Particle stiffness & $k_n$ & 0.1 \\ \hline
    Particle-particle damping & $k_d$ & $0.5 \sqrt{\frac{k_n m_i m_j}{ (m_i + m_j)}}$ \\ \hline
    Particle-wall stiffness & $k_w$ & 0.1 \\ \hline
    Particle-wall damping & $k_q$ & $0.5 \sqrt{k_w m_i}$ \\ \hline
    Friction coefficient & $\mu$ & $0.1$ \\ \hline
    \hline
  \end{tabular}
\end{center}
	 \caption{Table of model parameters.}
\end{table}

\subsubsection{Pore length parameters}
\label{app:porelength}
For the results shown in figure \ref{fig:porelength}, 12 simulations were run with different pore lengths $L$, namely $L = \{ 0.14, 0.70, 1.41, 2.11, 2.82, 3.52, 4.23, 4.93, 5.63, 6.34, 7.04, 7.75 \}$.

\subsubsection{Pore diameter parameters}
\label{app:porediameter}
For the results shown in figure \ref{fig:porediameter}, simulations with different pore diameter were run, the inlet area is scaled according to the pore diameter such that the ratio between inlet area and pore cross section is kept constant. For each pore diameter 10 simulations with different random seed were run, values of the pore diameters are $D = \{ 1.76, 2.11,  2.82, 3.52, 4.23, 5.28, 6.34, 7.04 \}$.

\bibliographystyle{ieeetr}
\bibliography{citations.bib}
\end{document}